
\input phyzzx

\catcode`\@=11
\paperfootline={\hss\iffrontpage\else\ifp@genum\tenrm
 -- \folio\ --\hss\fi\fi}
\def\titlestyle#1{\par\begingroup \titleparagraphs
 \iftwelv@\fourteenpoint\fourteenbf\else\twelvepoint\twelvebf\fi
 \noindent #1\par\endgroup }
\def\GENITEM#1;#2{\par \hangafter=0 \hangindent=#1
 \Textindent{#2}\ignorespaces}
\def\address#1{\par\kern 5pt\titlestyle{\twelvepoint\sl #1}}
\def\abstract{\par\dimen@=\prevdepth \hrule height\z@ \prevdepth=\dimen@
 \vskip\frontpageskip\centerline{\fourteencp Abstract}\vskip\headskip }
\newif\ifYUKAWA  \YUKAWAtrue
\font\elevenmib   =cmmib10 scaled\magstephalf   \skewchar\elevenmib='177
\def\YUKAWAmark{\hbox{\elevenmib
 Yukawa\hskip0.05cm Institute\hskip0.05cm Kyoto \hfill}}
\def\titlepage{\FRONTPAGE\papers\ifPhysRev\PH@SR@V\fi
 \ifYUKAWA\null\vskip-1.70cm\YUKAWAmark\vskip0.6cm\fi
 \ifp@bblock\p@bblock \else\hrule height\z@ \rel@x \fi }

\def\schapter#1{\par \penalty-300 \vskip\chapterskip
 \spacecheck\chapterminspace
 \chapterreset \titlestyle{\ifcn@@\S\ \chapterlabel.~\fi #1}
 \nobreak\vskip\headskip \penalty 30000
 {\pr@tect\wlog{\string\chapter\space \chapterlabel}} }

\def\ssection#1{\par \ifnum\lastpenalty=30000\else
 \penalty-200\vskip\sectionskip \spacecheck\sectionminspace\fi
 \gl@bal\advance\sectionnumber by 1
 {\pr@tect
 \xdef\sectionlabel{\ifcn@@ \chapterlabel.\fi
 \the\sectionstyle{\the\sectionnumber}}%
 \wlog{\string\section\space \sectionlabel}}%
 \noindent {\S \caps\thinspace\sectionlabel.~~#1}\par
 \nobreak\vskip\headskip \penalty 30000 }


\papers

\def\lkakko{\vbox{\vskip0.065cm\hbox{(}\vskip-0.065cm}}
\def\rkakko{\vbox{\vskip0.065cm\hbox{)}\vskip-0.065cm}}
\def\YUKAWAHALL{\hbox to \hsize
 {\hfil \lkakko\twelvebf YUKAWA HALL\rkakko\hfil}}


\def\Endline{\hfil\break}


\def\addeqno{\ifnum\equanumber<0 \global\advance\equanumber by -1
 \else \global\advance\equanumber by 1\fi}


\mathchardef\Lag="724C
\def\sqr#1#2{{\vcenter{\hrule height.#2pt
 \hbox{\vrule width.#2pt height#1pt \kern#1pt\vrule width.#2pt}
 \hrule height.#2pt}}}


\def\cref#1{\rlap,\attach{#1)}}
\def\ref#1{\attach{#1)}}



\newdimen\ex@
\ex@.2326ex
\def\boxed#1{\setbox\z@\hbox{$\displaystyle{#1}$}\hbox{\lower.4\ex@
 \hbox{\lower3\ex@\hbox{\lower\dp\z@\hbox{\vbox{\hrule height.4\ex@
 \hbox{\vrule width.4\ex@\hskip3\ex@\vbox{\vskip3\ex@\box\z@\vskip3\ex@}%
 \hskip3\ex@\vrule width.4\ex@}\hrule height.4\ex@}}}}}}
\def\txtboxed#1{\setbox\z@\hbox{{#1}}\hbox{\lower.4\ex@
 \hbox{\lower3\ex@\hbox{\lower\dp\z@\hbox{\vbox{\hrule height.4\ex@
 \hbox{\vrule width.4\ex@\hskip3\ex@\vbox{\vskip3\ex@\box\z@\vskip3\ex@}%
 \hskip3\ex@\vrule width.4\ex@}\hrule height.4\ex@}}}}}}
\newdimen\exx@
\exx@.1ex
\def\thinboxed#1{\setbox\z@\hbox{$\displaystyle{#1}$}\hbox{\lower.4\exx@
 \hbox{\lower3\exx@\hbox{\lower\dp\z@\hbox{\vbox{\hrule height.4\exx@
 \hbox{\vrule width.4\exx@\hskip3\exx@%
 \vbox{\vskip3\ex@\box\z@\vskip3\exx@}%
 \hskip3\exx@\vrule width.4\exx@}\hrule height.4\exx@}}}}}}

\chardef\fontD="1A

\catcode`@=12

\pubnum={YITP/K-984}
\date={August 1992}

\titlepage
\title{Quantum Groups, $q$-Oscillators and Covariant Algebras}

\author{P.P. Kulish \foot{
Address after September 1992 : Steklov Mathematical
Institute, Fontanka 27, St.Petersburg 191011, Russia}}

\address{\null\hskip-8mm
Yukawa Institute for Theoretical Physics\break
Kyoto University,~Kyoto 606,~Japan}

\vskip3mm
\centerline{
-- Dedicated to the memory of M.C. Polivanov --}

\abstract{
The physical interpretation of the main notions of the quantum group
theory (coproduct, representations and corepresentations, action and
coaction) is discussed using the simplest examples of $q$-deformed
objects
(quantum group $F_q(GL(2))$, quantum algebra $sl_q(2)$, $q$-oscillator
and $F_q$-covariant algebra.)
Appropriate reductions of the covariant algebra of second rank
$q$-tensors give rise to the algebras of the $q$-oscillator and
the $q$-sphere.
A special covariant algebra related to the reflection equation
corresponds to the braid group in a space with nontrivial topology.
}

\chapter{Introduction}
The formalism of the quantum inverse scattering method (the R-matrix
approach),
which was used in [1] in order to formulate the theory of quantum Lie
groups and Lie algebras, contributed much to the growth of attention
towards these new mathematical objects and to the use of the latter by
theoretical physicists.

Convincing examples of the description of the symmetry properties of
physical models by means of quantum groups and algebras have been
constructed (see, e.g. [2-6]), both in the standard way, when
Hamiltonian of the model H commutes with the quantum algebra generators
$X_i: [H, X_i] = 0$, and in the more complicated situation of
conformal field theory.
Quantum groups and algebras are also used as a basis for the new
approaches to the possible structure of space-time on small distances
(see [7,8]), which gives us an opportunity to define naturally
quantum homogeneous spaces.

However, despite intensive development of the mathematical theory of
quantum groups (cf.,[9]), the physical interpretation of many
statements
in this field deserves, from our point of view, more attention.
Naturally, it is much easier to keep the discussion to simple examples.
For this purpose we will limit ourselves to the well-known quantum
group
$F_q(GL(2))$, the quantum algebra $sl_q(2)$, the $q$-deformed
oscillator and
(a new example, [10]) the simplest quadratic algebra {\cal K}, which is
defined
by the reflection equation and is covariant with respect to the quantum
group $F_q(GL(2))$.
The notions discussed, include coproduct, representations and
corepresentations, action and coaction.

Being the associative algebra, the quantum group $F_q(GL(2))$ is
defined
by four generators $a,b,c,d$, which satisfy the relations
$$\eqalign{
ab=qba,~ ac=qca,~ [a,d]=\omega bc  \ , \cr
bd=qdb,~ cd=qdc,~ [b,c]=0  \ , \cr
}
\eqno{(1)}$$
where $q$ is the complex deformation parameter and $\omega =q-q^{-1}$.
The relations (1) define multiplication in algebra $F_q$ and
can be written in a compact matrix form [1] by means of
the $2 \times 2$ matrix $T$
$$ T= \pmatrix{ a  &  b  \cr
               c  &  d  \cr
}
\eqno{(2)}$$
and a $4 \times 4$ matrix $R$ with diagonal $(q,1,1,q)$ and the only
under-diagonal element not equal to zero $R_{21,12}=\omega $,
$$ RT_1T_2 = T_2T_1R
\eqno{(3)}$$
where $T_1 = T \otimes I$ and $T_2 = I \otimes T$.
Rows and columns of the $R$-matrix in ${\cal C}^2 \otimes {\cal C}^2$
are numerated by a pair of indexes, for example 11, 12, 21, 22.
As a mathematical object, $F_q$ is a Hopf algebra.
This means the existence of three more operations along with
multiplication $\mu $ : a coproduct
$\Delta : F_q \rightarrow F_q \otimes F_q$, an antipode
$s: F_q \rightarrow F_q$
and a counit $\epsilon : F_q  \rightarrow {\cal C}$.
According to the condition that $\Delta $ and $\epsilon $ are
homophormisms and
$s$ is an antihomophormism ($s(ab) = s(b)s(a)$), these three maps are
defined on generators and extended to the whole algebra $F_q$.
It is convenient to write them in matrix form [1]
$$\eqalign{
\Delta (T)=T (\otimes) T,~~ \Delta (t_{ij})= \sum_kt_{ik} \otimes
t_{kj}  \cr
s(T)T = Ts(T) = I,~~ \epsilon (T) = I  \ ,  \cr
}
\eqno{(4)}$$
where, as above, $I$ is the unit $2 \times 2$ matrix.
The operations $\mu ,\Delta ,\epsilon ,s$ are connected by a set of
axioms [1,11,15]
the validity of which is almost obvious in the $R$-matrix approach.

The quantum algebra $sl_q(2)$ is defined by three generators
$J,X_+,X_-$, which satisfy commutation relations
$$\eqalign{
[J,X_{\pm}] = \pm X_{\pm} \  ,  \cr
[X_+,X_-] = (q^{2J} - q^{-2J})/(q-q^{-1}) = [2J]_q \  , \cr
}
\eqno{(5)}$$
where the notation $[x]_q = (q^x-q^{-x})/\omega $ has been introduced.
The three additional maps, which define the structure of Hopf algebra
(quasitriangular one [11]) are fixed by the relations
$$\eqalign{
\Delta (J)=J \otimes 1 + 1 \otimes J \  ,
\Delta (X_{\pm})= X_{\pm} \otimes q^{-J} + q^J \otimes X_{\pm} \  , \cr
s(J)=-J, ~~s(X_{\pm})=-q^{\mp1}X_{\pm} \  ,  \cr
\epsilon (J)=\epsilon (X_{\pm}) = 0, ~~\epsilon (1)=1  \cr
}
\eqno{(6)}$$
Relations (5),(6) can also be written in the $R$-matrix form using
upper $+$ and low $-$ triangular matrices $L^{\pm}$[1,12].

The associative algebra ${\cal A}(q)$ of the deformed oscillator can be
defined even more easily : it has, like $sl_q(2)$, three generators
$A,A^+,N$ which satisfy
$$\eqalign{
[N,A] = -A,~~[N,A^+] = A^+ \  , \cr
[A,A^+] = q^{-2N} \  ,  \cr
}
\eqno{(7)}$$
but the corresponding Hopf algebra structure is not known.

Finally the associative algebra {\cal K}, connected with the reflection
equation [10], is defined by four generators $\alpha ,\beta ,\gamma
,\delta $, which
satisfy quadratic relations $(\omega  = q-q^{-1})$
$$\eqalign{
[\alpha ,\beta ]= \omega \alpha \gamma , ~~\alpha \gamma =q^2\gamma
\alpha ,~~ [\alpha ,\delta ]=\omega (q\beta +\gamma )\gamma  \  , \cr
[\beta ,\gamma ]=0,~~ [\beta \delta ] = \omega \gamma \delta ,~~ \gamma
\delta =q^2\delta \gamma  \  .  \cr
}
\eqno{(8)}$$
The Hopf algebra structure for this is not known either.
The $R$-matrix form of (8) uses a $2 \times 2$ matrix K of generators
$\alpha ,\beta ,\gamma ,\delta $ and the $R$-matrix of the quantum
group $F_q(GL(2))$ (3):
$$RK_1R^{t_1}K_2 = K_2R^{t_1}K_1R
\eqno{(9)}$$
where $R^t_1$ is a matrix transposed in the first space of
${\cal C}^2 \otimes {\cal C}^2$.
Thus the diagonal in $R^{t_1}$ is the same as in the R, and the only
non zero non diagonal element is situated in the right upper corner
of the matrix.

If we decide to use the above mentioned algebras as algebras of
observables, it is necessary to introduce real structure
( *- operation ) into them.
For example from $sl_q(2,{\cal C})$ one can get
$su_q(2),~su_q(1,1),~sl_q(2,{\cal R})$ with suitable restrictions
on the deformation parameter $(q \in {\cal R}~ {\rm and}~ |q| = 1~
{\rm representively})$.
However we will not discuss the real forms in this paper.

We should point out that all the above mentioned associative algebras
$F_q(GL(2)), \Endline
sl_q(2),~ {\cal A}(q)~ {\rm and}~ {\cal K}$ have central
elements
$$\eqalign{
det_qT = & ad-qbc;~~c_2 = X_-X_+ + [J]_q[J+1]_q  \  ; \cr
c_2(q) = & A^+A - [N;q^{-2}]~~;  \cr
z_1= & \beta -q\gamma ,~~z_2=\alpha \delta -q^2\beta \gamma  \  , \cr
}
\eqno{(10)}$$
where the notation $[n ; q] = (q^n-1)/(q-1)$ is also used.

The main aim of the paper is to point out close connections between
these algebras, some of which are well known, for example the duality
of
$sl_q(2)$ and $F_q(SL(2))$ [1], and to suggest the physical
interpretation of these connections.


\chapter{Connections between quantum algebras}

The deformed oscillator algebra ${\cal A}(q)$ can be obtained from
$sl_q(2)$ by means of the contraction [13,14]:
$$\eqalign{
A = \lim_{f \rightarrow 0} f\omega ^{1/2}X_+,~~
A^+ =\lim_{f \rightarrow 0} f\omega ^{1/2}X_-   \cr
q^{-N}=\lim_{f \rightarrow 0}fq^J  \   ,   \cr
}
\eqno{(11)}$$
and for the Casimir operator (10) one gets
$\lim f^2\omega c_2 = c_2(q) + q^2/(q^2-1)$.
In particular relations (6),
which define the Hopf algebra structure of $sl_q(2)$,
do not survive in this limit.
However, it is possible to obtain, a finite expression in the right
hand side of (6) for coproduct $\Delta $,
$$\eqalign{
\psi (N)  = & N \otimes I - I \otimes J ~~, \cr
\psi (A)  = & A \otimes q^{-J}+\omega ^{1/2}q^{-N} \otimes X_+ ~~,\cr
\psi (A^+)= & A^+ \otimes q^{-J} + \omega ^{1/2}q^{-N} \otimes X_- ~~,
\cr
}
\eqno{(12)}$$
and to interpret it as a map $\psi$ from algebra ${\cal A}(q)$ into
the tensor product ${\cal A}(q) \otimes sl_q$ (2).
The latter preserves the commutation relations (7) of algebra
${\cal A}(q)$, for example,
$$[\psi (A), \psi (A^+)] = q^{-2\psi (N)} \  .
\eqno{(13)}$$
This map, which satisfies the properties of consistency with the
coproduct:
$(\psi \otimes id) \circ \psi = (id \otimes \Delta ) \circ \psi$
and counit $(id \otimes \epsilon ) \circ \psi = id$ of the quantum
algebra
$sl_q(2)$, is called the coaction of $sl_q(2)$ on ${\cal A}(q)$,
and the algebra
${\cal A}(q)$ is called the $sl_q(2)$-comodule algebra [15].
While the coproduct makes it possible to define the action of the Hopf
algebra in the tensor product of its representations, the coaction,
in this case
$\psi$, defines the representation of the algebra ${\cal A}(q)$ in the
tensor
product of the representation of ${\cal A}(q)$ with a representation of
$sl_q(2)$.
Thus, with the coproduct for $sl_q(2)$ interpreted as an addition of
$q$-spins, the coaction $\psi$ can be interpreted as a kind of
addition
of $q$-oscillator with $q$-spin, resulting in the reproducing of the
$q$-oscillator algebra.

If we limit ourselves to the Fock representation of ${\cal A}(q)$
(the other representations are obtained in [13]),
which is ${\cal H}_F$
with basic vectors :
$|0\rangle ({\rm such~ that} A|0\rangle = 0)~ {\rm and}~
 |n\rangle \simeq (A^+)^n |0\rangle$
and choose the Hamiltonian of the $q$-oscillator to be
$$\eqalign{
     H = &  A^+A = [N;q^{-2}] = (1-q^{-2N})/(1-q^{-2})\  ,\cr
spec H = &  [n;q^{-2}],~~n = 0,1,2, \dots  \cr
}
\eqno{(14)}$$
then the Hamiltonian $H_I$ of the interacting in a special way
$q$-oscillator and $q$-spin will have the same structure of spectrum
(with additional multiplicities) :
$spec H_I = \{[n;q^{-2}],\nu _n(j),n= -j, -j+1, \dots \}$, where $j$ is
the value of the $q$-spin and $\nu _n(j)$ correspondent multiplicity
$\nu _n(j) = n+j+1,~ n\in -j, -j+1, \dots, j ;~ \nu _n(j) =2j+1,~ n >
j$.
The corresponding expression for the interacting Hamiltonian
can be obtained by introducing the right hand side of (12) into (14)
$H_I = \psi (A^+)\psi (A)$
$$H_I = A^+Aq^{-2J} + \omega q^{-2N}X_-X_+ + \omega
^{1/2}q^{-J-N}(A^+X_+ + AX_-)
\eqno{(15)}$$
The Hamiltonian (15) acts in the ${\cal H}_F \otimes V_j$ space, where
$V_j$ is an irreducible representation of $sl_q(2)$ with
$dim V_j=2j+1$.
This space is decomposed into a direct sum of $2j+1$ irreducible Fock
space representations of the $q$-oscillator algebra ${\cal A}(q)$.
Due to the consistency condition of the coaction $\psi $ with coproduct
$\Delta $ of $sl_q(2)$, it is possible to consider the interaction of
the
$q$-oscillator with any number of $q$-spins whilst preserving the
spectrum
structure of the Hamiltonian (14).
In frame of the $q$-Clebsch-Gordan-Wigner-Racah calculus it corresponds
to the contraction procedure w.r.t. one of the $q$-spin variables.

The algebra {\cal K} (8) and the quantum group $F_q(GL(2))$ are
connected
in the same way [10].
The map (coaction)
$\varphi:{\cal K} \rightarrow F_q \otimes {\cal K}$ is given on the
generators $\alpha ,\beta ,\gamma ,\delta $ by matrix multiplication :
$$ \varphi:K \rightarrow K_T = TKT^t
\eqno{(16)}$$
for example $\varphi(\beta ) =ac\alpha  + bc\gamma  + ad\beta  +
bd\delta $.
As in the previous case (12) the coaction (16) is consistent with the
comultiplication $\Delta $ and counit $\epsilon $ of $F_q$.
Like $\psi$, it also preserves the commutation relations (8) for the
transformed generators $\varphi(\alpha ), \varphi(\beta ), \dots$.
This connection between the algebras {\cal K} and $F_q$ enables us to
characterize {\cal K} as $F_q$-comodule algebra [15].
Thus, if $F_q$ and ${\cal K}$ are considered as algebras of
observables, the
existence of the coaction $\varphi$ makes it possible to reproduce
the structure of {\cal K} in the combined system
$F_q \otimes {\cal K}$ and
with fixed representations $V$ and $W$ for $F_q$ and ${\cal K}$ to
look for the
expansion of $V \otimes W$ on irreducible representations of algebra
${\cal K}$.

In a Hopf algebra, e.g. $F_q(GL(2))$, there are two main maps :
multiplication $\mu $ and comultiplication $\Delta $.
Hence one can consider representations of $F_q$ as an associative
algebra
by appropriate linear operators in a linear space.
Then $\mu $ corresponds to multiplication of the operators and the
existence of
the comultiplication $\Delta $ in $F_q$ defines a representation of
$F_q$ in
the tensor product of given representations $V_1 \otimes V_2$
(cf.[17]).
One can consider also corepresentations of $F_q$ when the defined map
(comultiplication) $\Delta $ is consistent with the corepresentation
map
$\varphi$ (coaction).
For example, the space $V$ with basic elements $x,y$ is a
corepresentation for $F_q(GL(2))$ with coaction
$\varphi:V \rightarrow F_q \otimes V$ given by
$$ \varphi(x) = ax + by,~~\varphi(y) = cx + dy $$
for these maps $\varphi, \Delta ~ {\rm and}~ \epsilon $ are consistent
$$ (\Delta  \otimes id) \circ \varphi = (id \otimes \varphi) \circ
\varphi,~~
(\epsilon  \otimes id) \circ \varphi = id
\eqno{(17)}$$
Provided the elements $x,y$ commute with the generators
$a,b,c,d$ of $F_q$ one can show that the relation $xy = qyx$
(the quantum plane [18]) is preserved under the coaction :
$\varphi(x)\varphi(y) = q \varphi(y)\varphi(x)$.
Thus $V$ can be characterized as an $F_q$-comodule algebra.
As in the previous example of the $q$-oscillator and $q$-spin one
can interpret corepresentations as addition of two physical systems,
considering $F_q$ and $V$ as some algebras of observables, such
that in the
combined system $F_q \otimes V$ one can reproduce by the coaction map
$\varphi$ one of the original algebras i.e. $V$.

By analogy with representations we can take few copies of
corepresentations e.g. $V_1, V_2$.
However, to have a corepresentation in the tensor product
$V_1 \otimes V_2$ the generators $x_1, y_1$ of $V_1$ must not
commute with those $x_2,y_2$ of $V_2$.
In order to define corepresentation they should satisfy the relations
$(\omega  = q-q^{-1})$
$$\eqalign{
x_iy_i = & qy_ix_i,~~~x_1y_2 = y_2x_1 + \omega x_2y_1  \cr
x_2y_1 = & y_1x_2,~~~x_1x_2 = qx_2x_1,~~~y_1y_2 = qy_2y_1   \cr
}
\eqno{(18)}$$

The simplest way to explain the covariance of the relations
(18) is the $R$-matrix
formalism [1] and the exchange or Zamolodchikov - Faddeev algebra
[13b].
Using the $R$-matrix (3) and two component spinors
$v_i^t = (x_i,y_i), i = 1,2$ one can write (18) in the compact form
$${\hat R} v_i \otimes v_i = qv_i \otimes v_i,~~
{\hat R} v_2 \otimes v_1 = v_1 \otimes v_2 ,~~
{\hat R} = {\cal P}R\  .
\eqno{(19)}$$
It is obvious that due to (3) these relations are invariant w.r.t.
the coaction $v_i \rightarrow \varphi(v_i) = Tv_i$.
The element $x_1y_2 - qy_1x_2 = y_2x_1 - q^{-1}x_2y_1$ is central.
In fact, relations (18) coincide with (1) and the central element with
$det_q$.
However, one can extend this composition rule to any number of
corepresentations $V_j, j=1,2,3, \dots$ with relations (19) for any
ordered pair of indices $i > j$.

This extension procedure can be applied also to the covariant algebra
{\cal K} (8).
However we shall do it using the equivalent algebra {\cal B} [2,10,16],
directly related to the braid group.
The algebra {\cal B} has a $2 \times 2$ matrix of generators $M$
satisfying
the following reflection equation
$$M_1R^{-1}M_2{\tilde R}^{-1} = R^{-1}M_2{\tilde R}^{-1}M_1.
\eqno{(20)}$$
where ${\tilde R} = {\cal P}R{\cal P}~ {\rm and}~ {\cal P}$ is
the permutation (flip)
operator of two spaces ${\cal C}^2 \otimes {\cal C}^2$.
The corresponding coaction of the quantum group $F_q(GL(2))$ is
$$M \rightarrow M_T = TMT^{-1}
\eqno{(21)}$$
Let us consider two copies of this algebra ${\cal B}_i, i = 1,2,$ with
matrices of generators $M(i)$ satisfying (20).
In order for the product of these matrices $M(1)M(2) = M^{(2)}$ to
satisfy (20) it is sufficient to require the following
relations among the generators
of ${\cal B}_1$ and ${\cal B}_2$
$$ M_1(2)R^{-1}M_2(1)R = R^{-1}M_2(1)RM_1(2) \  .
\eqno{(22)}$$
The proof is straightforward, however, we should point out probably
more
obvious graphical proof which follows from the connection of the
reflection equation (20) with the braid group in a solid handlebody
[6].
Multiplying (20) by {\cal P} from left and right, identifying
$M_2 = \tau , {\cal P}R^{-1}= \sigma _1$ one obtains the additional
relation
[6,19]
between generators $\tau $ and $\sigma _1$ of $B_n^{(1)}$ (cf. Fig 1)
$$ \tau \sigma _1\tau \sigma _1 = \sigma _1\tau \sigma _1\tau \  .
\eqno{(23)}$$
If there are two handles in the body the braid group $B_n^{(2)}$ has
two additional generators $\tau _i, i=1,2$ which satisfy (23) and
relation between them (cf. Fig.2)
$$\tau _2\sigma _1\tau _1\sigma ^{-1}_1 = \sigma _1\tau _1\sigma
_1^{-1}\tau _2 \   .
\eqno{(24)}$$
which coincides with (22).
Composition of the generators $\tau _1\tau _2$ corresponds to M(1)M(2)
and
satisfies the same relation (23), (20) as one generator (Fig.3).
One can develop this for any number of handles $g,B_n^{(g)}$.
The algebraic properties of constructed set of algebras
${\cal B}_i, i=1,2, \dots, g$ (central elements, abelian
subalgebras, representations) are discussed in [22].

One can also interpret some quantum algebras as spaces of quantum
states of some physical model.
The covariant algebra {\cal K} which due to its transformation property
$$ k_{ij} \rightarrow \sum_{m,n}T_{im}T_{jn}k_{mn}
\eqno{(25)}$$
may be called a $q$-tensor of the second rank (generalizations to
higher
ranks were proposed in [10]), has two central elements (10).
These central elements enter into the characteristic equation for the
matrix {\cal K} [10] :
$$K\epsilon _qK = z_2\epsilon _q - qz_1K ~~,
\eqno{(26)}$$
where $\epsilon _q$ is the $2 \times 2$ off diagonal matrix of the
quantum
metric for $F_q(GL(2))$.
The algebra {\cal K} is reduced by fixing one of central elements e.g.
$z_1 =\beta -q\gamma  = 0$.
The reduced or factor algebra ${\cal L}_q$ is generated by three
elements $\alpha ,\gamma ,\delta  ~~(\beta =q\gamma )$ satisfying
relations
$$ \alpha \gamma =q^2\gamma \alpha ,~ \gamma \delta =q^2\delta \gamma
,~ [\alpha ,\delta ]=q(q^2-q^{-2})\gamma ^2 \   .
\eqno{(27)}$$
These are the relations of the generators of the quantum two
dimensional
sphere [20].
The second central element $z_2=\alpha \delta -q^3\gamma ^2$
corresponds to the radius
of the sphere and the matrix $K\epsilon _q$ corresponds to the
classical matrix
$S=\Sigma x_i\sigma _i$ with the constraint $S^2 = \Sigma x^2_i$
(cf.(26)).
Like $L_2(S^2)$ this algebra ${\cal L}_q$ being covariant w.r.t.
the quantum group $F_q(SU(2))$ coaction
$e_i \rightarrow \Sigma t_{ij}^{(1)} \otimes e_j$ is decomposed as a
linear
space into a direct sum of irreducible (co)representations
${\cal L}_q=\Sigma V_n, n=0,1,2, \dots$.
There exists an invariant integral $\nu $ on ${\cal L}_q$ as a linear
functional $\nu (\eta ) \in {\cal C},~ \eta  \in {\cal L}_q$.
Using this invariant integral one can think on ${\cal L}_q$ as a
space of states of the quantum particle on the $q$-deformed sphere.
The $q$-deformed statistics of these particles is discussed in [21].
Being dual to $F_q (SU(2))$ the quantum algebra $su_q(2)$
acts on ${\cal L}_q$ by ``differential" or difference operators and
its central element $c_2$ (10) is the $q$-Casimir operator in
${\cal L}_q$, which has
eigenvalue $[n]_q[n+1]_q$ over each irreducible component $V_n$.

By another identification of the reduced algebra {\cal K} generators
$$\alpha  \rightarrow A^+,~~ \delta  \rightarrow -A,~~
q^3(1-q^{-4})\gamma ^2 \rightarrow q^{-2N},~~ q \rightarrow q^{1/2}  $$
one gets the $q$-oscillator algebra (7).

\chapter{Conclusion}

The preceding considerations give quite different physical
interpretations of the $q$-deformed algebras :
as the symmetry algebra of a model, as an algebra of observables and
even as a space
of physical states.
After appropriate quasiclassical limits all these interpretations have
corresponding classical counterparts.
Hence the $q$-deformed objects are analogous
with respect to many properties
to the usual Lie groups and Lie algebras though their use
and interpretation is sometimes rather elaborated.

In attempting to understand these ideas more completely
we shall sadly miss the contributions of the late
Professor M.C. Polivanov.

\vskip20mm
\centerline{\fourteenbf References}
\vskip5mm

\item{1.} Faddeev L., Reshetikhin N. and Takhtajan L.
Algebra Analiz {\bf 1}(1989) 178; \Endline
Leningrad Math J. 1(1990) 193.
\item{2.} Moore G. and Reshetikhin N.~~ Nucl.Phys. {\bf B328},
(1989)p.557
\item{3.} Alvarer-Gaum\'e L., G\'omez C. and Sierra G. ~~Nucl.Phys.
B330, 347(1990)
\item{4.} Pasquier V. and Saleur H.~~Nucl.Phys. {\bf B330}, 523(1990);
Jimbo M., Miki K., Miwa T. and Nakayashiki A. ~~Correlation functions
of the XXZ model for $\Delta <-1$.
Preprint RIMS-877(1992)
\item{5.} Kulish P. and Sklyanin E. ~~J.Phys {\bf A24}, L435(1991)
\item{6.} Kulish P. ~~Quantum groups and quantum algebras as symmetries
of dynamical systems, Preprint YITP/K-959, Kyoto (1991)
\item{7.} Zumino B. Introduction to differential geometry of quantum
groups.~~ Preprint UCB-PTH-62/91, Berkeley (1991)
\item{8.} Carow-Watamura V. et al ~~Z.Phys. {\bf C48}, 159 (1990)
\item{9.} Kulish P.P.(Ed.)  Proc.Euler Inter.Math.Inst.
``Quantum Groups", Lect Notes Math. v. {\bf 1510} (Springer,
Berlin)1992,
398pp
\item{10.} Kulish P. and Sklyanin E. ~~Algebraic structures related to
reflection equations, ~Preprint YITP/K-980, Kyoto (1992) ;
J.Phys. {\bf A25} (1992).
\item{11.} Drinfeld V.G. ~Quantum groups. Proc. ICM-86, 1 (Berkeley)
1987, p.798
\item{12.} Takhtajan J.A. Quantum groups, Lect.Notes Phys. {\bf 370},
3 (1990)
\item{13.} Kulish P.P. Teor.Matem.Fiz. {\bf 86}, 157(1991);
Phys.Lett. {\bf A161} (1991)51
\item{14.} Chaichian M. and Kulish P. ~~Phys.Lett. {\bf B234}(1990)72
\item{15.} Majid S. Int.J.Mod.Phys. {\bf A5}, 1(1990)
\item{16.} Majid S. In:[9], p.79 ; Jour.Math.Phys. {\bf 32}(1991)3246 ;
Preprint DAMTP/92-12
\item{17.} Faddeev L.D. and Takhtajan L.A. ~~Lect.Notes Phys.
{\bf 246}. 16
(1989)
\item{18.} Manin Yu.I.~~Topics in noncommutative geometry
Princeton (1990)
\item{19.} Sossinsky A.B. ~~In:[9], p.354
\item{20.} Podles P. ~~Lett.Math.Phys. {\bf 14}, 193 (1987) ;
Noumi M. and Mimachi K. \Endline
Comm.Math.Phys. {\bf 128}, 521 (1990)
\item{21.} Podles P.~~Quantization enforces interaction.
Preprint RIMS-817, Kyoto (1991)
\item{22.} Kulish P. and Sasaki R.~~Covariant properties of reflection
equation algebras. ~~Preprint YITP/U-92-21 (1992)

\endpage
\bye